\documentclass[12pt]{article}
\usepackage[margin=1.1in]{geometry}
\usepackage{float}

\usepackage{enumerate}
\usepackage{amsmath}
\usepackage{amssymb}
\usepackage{amsthm}
\usepackage{color}
\usepackage{xcolor}
\usepackage{graphicx}

\usepackage{bbm}
\usepackage{color}
\usepackage{geometry}
\usepackage{bbding}
\usepackage{ dsfont }
\usepackage{caption}
\usepackage{stackrel}
\usepackage{graphicx,xcolor}
\usepackage{tikz}
\usepackage{hyperref}
\usepackage{enumerate}
\usepackage{multicol}
\usepackage{soul}
\usepackage{tabularx}
\usepackage{multirow}
\usepackage{mdwlist}
\usepackage{enumitem}
\usepackage{accents}
\usepackage{mathtools}
\usepackage{subcaption}

\makeatletter
\DeclareRobustCommand\widecheck[1]{{\mathpalette\@widecheck{#1}}}
\def\@widecheck#1#2{%
    \setbox\z@\hbox{\m@th$#1#2$}%
    \setbox\tw@\hbox{\m@th$#1%
       \widehat{%
          \vrule\@width\z@\@height\ht\z@
          \vrule\@height\z@\@width\wd\z@}$}%
    \dp\tw@-\ht\z@
    \@tempdima\ht\z@ \advance\@tempdima2\ht\tw@ \divide\@tempdima\thr@@
    \setbox\tw@\hbox{%
       \raise\@tempdima\hbox{\scalebox{1}[-1]{\lower\@tempdima\box
\tw@}}}%
    {\ooalign{\box\tw@ \cr \box\z@}}}
\makeatother

\usepackage{threeparttable}
\newcommand{\deleq}{\stackrel{\Delta}{=}}
\theoremstyle{empty}
\newtheorem{example}{Example}
\newtheorem{proposition}{Proposition}
\newtheorem{theorem}{Theorem}

\newtheorem{lemma}{Lemma}

\usepackage{mathtools}


\newcommand{\R}{R}
\renewcommand{\r}{r}
\newcommand{\rone}{\r_1}
\newcommand{\rtwo}{\r_2}

\newcommand{\I}{1}
\newcommand{\D}{D}
\newcommand{\W}{\Omega}
\newcommand{\U}{2}

\newcommand{\qIwe}{q^{1*}}
\newcommand{\qUwe}{q^{2*}}

\newcommand{\qI}{q^{1}}
\newcommand{\qU}{q^{2}}
\newcommand{\q}{q}
\newcommand{\betaIs}{\beta^{\s}}

\newcommand{\Pwe}{\P^{*}}
\newcommand{\lambdag}{\underline{\lambda}}
\newcommand{\lambddag}{\bar{\lambda}}
\newcommand{\gbar}{\bar{g}}
\newcommand{\Lambthree}{\Lambda_3}
\newcommand{\pbar}{\bar{p}}
\newcommand{\Q}{Q}
\renewcommand{\a}{\mathbf{a}}
\newcommand{\n}{\mathbf{n}}
\newcommand{\w}{\omega}

\renewcommand{\S}{S}
\newcommand{\s}{s}

\newcommand{\piwe}{\pi^{*}}
\newcommand{\qwe}{q^{*}}
\newcommand{\lamb}{\lambda}
\renewcommand{\(}{\left(}
\renewcommand{\)}{\right)}
\newcommand{\f}{f}
\newcommand{\fone}{\f_1}
\newcommand{\ftwo}{\f_2}
\newcommand{\fr}{f_r}

\newcommand{\aonea}{\alpha_1^\a}
\newcommand{\aonen}{\alpha_1^\n}
\newcommand{\atwo}{\alpha_2}
\newcommand{\bone}{b_1}
\newcommand{\btwo}{b_2}

\renewcommand{\(}{\left(}
\renewcommand{\)}{\right)}

\newcommand{\fwe}{f^{*}}

\newcommand{\pro}{\mathrm{Pr}}
\newcommand{\aone}{\alpha_1}
\newcommand{\lambbar}{g}
\newcommand{\Lambone}{\Lambda_1}
\newcommand{\Pione}{\Pi^1}
\newcommand{\Pitwo}{\Pi^2}
\newcommand{\Lambtwo}{\Lambda_2}
\renewcommand{\P}{\mathrm{P}}

\newcommand{\conew}{c_1^{\w}}
\newcommand{\ctwo}{c_2}
\newcommand{\betaIa}{\beta^{\a}}
\newcommand{\betaIn}{\beta^{\n}}
\begin{document}

\title{Information Design for Regulating Traffic Flows under Uncertain Network State}
\author{Manxi Wu, and Saurabh Amin
\thanks{M. Wu is with the Institute for Data, Systems, and Society, S. Amin
is with the Department of Civil and Environmental Engineering, and Institute for Data, Systems and Society, Massachusetts Institute of Technology, 77 Massachusetts Ave., Cambridge, MA. USA, \{manxiwu, amins\}@mit.edu}%
}
\date{}
\maketitle
\begin{abstract}
Traffic navigation services have gained widespread adoption in recent years. The route recommendations generated by these services often leads to severe congestion on urban streets, raising concerns from neighboring residents and city authorities. This paper is motivated by the question: How can a transportation authority design an information structure to induce a preferred equilibrium traffic flow pattern in uncertain network state conditions? We approach this question from a Bayesian persuasion viewpoint. We consider a basic routing game with two parallel routes and an uncertain state that affects the travel cost on one of the routes. The authority sends a noisy signal of the state to a given fraction of travelers. The information structure (i.e., distribution of signals in each state) chosen by the authority creates a heterogeneous information environment for the routing game. The solution concept governing the travelers' route choices is Bayesian Wardrop Equilibrium. We design an information structure to minimize the average traffic spillover -- the amount of equilibrium route flow exceeding a certain threshold -- on one of the routes. We provide an analytical characterization of the optimal information structure for any fraction of travelers receiving the signal. We find that it can achieve the minimum spillover so long as the fraction of travelers receiving the signal is larger than a threshold (smaller than 1).
\end{abstract}

\section{INTRODUCTION}
Today, travelers heavily rely on a variety of information sources in making their day-to-day travel decisions. These sources range from public systems (radio broadcasts, road message signs) to navigation apps installed on their mobile devices. One approach to study the impact of traffic information on route choices and/or departure time decisions of travelers (players) is to formulate and analyze a game-theoretic model under a fixed information environment, which specifies the knowledge of individual travelers about the uncertain network state (e.g., presence of incidents or capacity perturbations) and the equilibrium behavior of other travelers~\cite{arnott1991does}, \cite{acemoglu2018informational}, \cite{wu2018value}, \cite{khan2018bottleneck}. This paper is motivated by the need to extend this approach for designing an ``optimal" information structure that can be used to regulate traffic flows in a network with uncertain state. We adopt the viewpoint of Bayesian persuasion and focus on identifying the distribution of information signal (conditional on the state) to induce an equilibrium outcome that is close to a target flow pattern which reflects the preference of a central authority (information designer).

Practically, our setup is motivated by the several concerns raised by city authorities and residents of areas that have witnessed significant increase in traffic congestion in their neighborhood streets due to route recommendations provided by the navigation apps~\cite{bliss_bliss_2015}, \cite{reporter_newspapers_2016}, \cite{city_news_service_city_news_service_2017}, \cite{foderaro_2017}. In some cases, these routes pass through school areas, evacuation zones, construction sites, or active incidents. Increased traffic through these regions naturally raises noise and safety concerns. In other cases, recommended routes are often comprised of secondary streets that were not designed for heavy and prolonged rush hour traffic. As a result, these streets can witness deterioration in infrastructure condition, and further decrease in their traffic carrying capacity. To address these concerns, cities and transportation agencies -- henceforth  jointly referred as \emph{central authority} -- are now playing an active role in communicating their preferred limits on the usage of neighborhood streets to traffic information providers~\cite{geha_geha_2016}, \cite{barragan_2015}. This raises the question: \emph{How can the central authority reduce traffic spillover (average flow exceeding a specific threshold) on certain routes by designing the information environment faced by travelers?} 

We present a stylized model to address this question. Our model captures an important (and practically relevant) feature of traffic information design: not all travelers can receive the signal sent by the central authority. For example, some travelers may not have access to or choose not to use the information signal. We capture this feature in a Bayesian congestion game, where the heterogeneous information environment is determined not only by the the signal sent by the central authority, but also due to the amount of travelers with access to this signal. This game allows us to formulate and solve the information design problem, in which the distribution of information signal is chosen by the central authority to regulate the induced equilibrium traffic flows.

We consider a transportation network with two parallel routes, where one route is prone to a random capacity-reducing event (incident), and the other is not. Incident state results in increased travel cost on the first route. The cost of each route is an increasing (affine) function of the flow on that route. The central authority (information designer) knows the true state (i.e., whether or not the incident happened), and chooses an information structure that is used to send the signal to a fraction of travelers. Travelers are strategic in that they choose routes with the minimum expected cost based on their information of the state. The induced route flow is a Bayesian Wardrop equilibrium corresponding to the information structure chosen by central authority. The objective of the central authority is to minimize the average spillover (in Bayesian Wardrop equilibrium) on a pre-selected route beyond a threshold flow.

Our solution approach to optimal information design is based on two key steps: First, for any fraction and any (feasible) information structure, we characterize the unique equilibrium route flow of the Bayesian routing game. We find that all feasible information structures can be partitioned into two sets, each of which have qualitatively distinct impact on the equilibrium route flows (Proposition \ref{prop:BWE}). Second, based on the analysis of equilibrium route flows, we pose the problem of selecting the optimal information structure as an optimization problem, which is nonlinear and non-convex in the information structure. 

Still, we manage to analytically solve the abovementioned optimization problem. For any fraction of travelers receiving the information signal, we fully characterize the optimal information structure (Proposition \ref{prop:no_persuasion}, Theorem \ref{theorem:opt_persuasion}). We find that if when the probability of the incident state is low, it is optimal not to provide any state information to travelers, and the spillover is zero. On the other hand, when the probability of the incident state is high, the optimal information structure depends on the fraction of travelers who receive the signal. We show that given the optimal information structure, the spillover is positive only in the incident state. Moreover, we find that the minimum average spillover can be achieved by the optimal information structure as long as the fraction of travelers receiving the signals exceeds a certain threshold fraction. Interestingly, this threshold fraction is also the fraction that achieves the minimum equilibrium average cost of all travelers under the optimal information structure (Proposition \ref{eq_route_cost}).

Our work contributes to the recent literature of information design. The papers \cite{das2017reducing} and \cite{tavafoghi2017informational} focused on designing the optimal information structure that sends signals of uncertain state to \emph{all} travelers in order to minimize the overall traffic congestion (as opposed to regulating traffic flows on selected route(s)). The ideas used in our approach are related to the broader literature on Bayesian persuasion; see \cite{kamenica2018bayesian} for a comprehensive review. Other relevant literature includes Bayesian persuasion with multiple receivers \cite{bergemann2016bayes}, \cite{mathevet2017information}, and persuasion with private information \cite{kolotilin2017persuasion}. The distinguishing aspects of our paper are: (1) the objective of the central authority is to minimize the average traffic spillover on chosen route, instead of average travel time; (2) travelers' route choices are evaluated in Bayesian Wardrop equilibrium under the heterogenous information environment created by the information signal; (3) characterization of optimal information structure for \emph{any} fraction of travelers receiving the signal. 

Rest of the paper is organized as follows: In Sec. \ref{sec:model}, we present our model of Bayesian congestion game and formulate the information design problem. In Sec. \ref{sec:BWE}, we characterize equilibrium outcomes of the game under heterogeneous information environment created by the signal sent to a fraction of travelers. Next, in Sec. \ref{sec:persuasion}, we present the optimal information structure for any given spillover threshold and any fraction of travelers with access to information signal. We analyze the impact of optimal information structure on travelers' costs in Sec. \ref{sec:cost}. Due to limited space, we do not present detailed proofs and only discuss the key ideas behind our results. 
  
 
 

\section{Model}\label{sec:model}
\subsection{Network}
Consider a traffic routing problem in which a single origin-destination pair is connected by two parallel routes $\R=\{\rone, \rtwo\}$. Non-atomic travelers (players) with total demand of $\D$ make route choices in the network. 

The cost (travel time) of each route $\r \in \R$ is an affine increasing function of the route flow $\fr$. Moreover, the cost function of $\rone$ is affected by network state $\w \in \W \deleq \{\a, \n\}$, which is uncertain. The cost function of $\rtwo$ is state-independent. Particularly, the cost functions are given by:  
\begin{align*}
    \conew(\fone)&= \left\{
    \begin{array}{cc}
        \aonea \fone+\bone, &\quad \text{if $\w=\a$,}  \\
         \aonen \ftwo+\btwo, &\quad \text{if $\w=\n$,} 
    \end{array}
    \right.\\
    \ctwo(\ftwo)&=\atwo \ftwo +\btwo.
\end{align*}
The state is drawn from a prior probability distribution $\theta$, where $\theta(\a)=p$ and $\theta(\n)=1-p$. 

In this model, state $\a$ (``accident'') corresponds to an incident on $\rone$; and state $\n$ (``nominal") indicates no incident on $\rone$. Route $\rtwo$ is not prone to incidents. A practical interpretation is that state $\a$ reduces the capacity of $\rone$, and so it faces higher rate of congestion as $\f_1$ increases (relative to state $\n$), i.e. $\aonea>\aonen$. Additionally, state does not impact the cost when there is no traffic, i.e. free-flow travel time $\bone$ is not affected by the presence of incident. 

For ease of presentation, we assume that $\rone$ is ``shorter" than $\rtwo$ in that its free-flow travel time is smaller, i.e. $\bone<\btwo$. In addition, the rate of congestion in $\rone$ is smaller than $\rtwo$ in state $\n$, but larger in state $\a$, i.e. $\aonea>\atwo>\aonen$. To avoid triviality, we also assume that the demand is sufficiently high so that players do not take one route exclusively: \begin{align}\label{high_d}
    \D > \frac{\btwo-\bone}{\aonen}.
\end{align}




\subsection{Information Structure and Traffic Flow Regulation}
We introduce a central authority (city or 
transportation agency) who has complete knowledge of the realized network state. This authority is an ``information designer'' in that she has the ability to shape the players' information about the state by way of sending them a (noisy) signal $\s\in\S \deleq \{\a,\n\}$. For example, in the context of transportation systems, the authority can influence players' knowledge of route conditions through advanced traveler information systems and/or navigation apps. The authority chooses an \emph{information structure} $\pi = (\pi(\s|\w))_{\s \in\S, \w \in \W}$, where $\pi(\s|\w)$ is the probability of sending signal $\s$ when the state is $\w$. Let $\Pi$ be the set of feasible information structures satisfying the following constraints: 
\begin{subequations}\label{feasible_pi}
\begin{align}
    \pi(\s|\w)&\geq 0, \quad \forall \s \in \S, \text{ and } \forall \w \in \W, \label{positive}\\
    \sum_{\s \in \S} \pi(\s|\w)&=1, \quad \forall \w \in \W, \label{sum_one}\\
    \pi(\n|\n) &\geq \pi(\n|\a), \label{informativeness}
\end{align}
\end{subequations}
Constraints \eqref{positive} - \eqref{sum_one} ensure that $(\pi(\a|\w), \pi(\n|\w))$ is a feasible probability vector for any $\w \in \W$. Constraint \eqref{informativeness} ensures that the signal $\n$ is more likely to be sent in state $\n$ than in state $\a$. Constraints \eqref{sum_one} and \eqref{informativeness} also imply that signal $\a$ is more likely to be sent in state $\a$ than in state $\n$, i.e. $\pi(\a|\a) \geq \pi(\a|\n)$. We use \eqref{informativeness} to avoid duplication of equivalent information structures. Note that one information structure becomes another equivalent structure by switching the signals $\a$ and $\n$.

Furthermore, in practice, it is reasonable to expect that some players may not have access to or choose not to use the signal sent by the authority. We refer the mass of players who receive the signals as population 1, and the remaining players who do not have access to signals as population 2. In our problem of optimal information design (Sec. \ref{sec:persuasion}), the fraction of population 1, denoted $\lamb \in [0,1]$, is taken as an exogenous parameter. 

The routing strategy of population $\I$ is a function of the received signal, denoted $\qI=\left(\qI_r(\s)\right)_{\r \in \R, \s \in \S}$, where $\qI_r(\s)$ is the amount of population $\I$ players who take route $\r$ when receiving the signal $\s$. On the other hand, the routing strategy of population $\U$ can be simply expressed by $\qU=\left(\qU_r\right)_{\r \in \R}$, where $\qU_r$ is the amount of population $\U$ players taking route $\r$. A strategy profile $\q\deleq (\qI, \qU)$ is feasible if it satisfies the following constraints: 
\begin{subequations}\label{feasible_q}
\begin{align}
    &\sum_{\r \in \R} \qI_r(\s) =\lamb \D, \quad \forall \s \in \S,\\
    &\sum_{\r \in \R} \qU_r = (1-\lamb)\D, \\
    &\qI_\r(\s)\geq 0, \quad \qU_r \geq 0, \quad \forall \r \in \R, \quad \forall \s \in \S.
\end{align}
\end{subequations}
We denote the feasible strategy set as $\Q$. The vector of flow is $\f=\left(\fr(\s)\right)_{\r \in \R, \s \in \S}$, where $\fr(\s)$ is the aggregate flow on route $\r$ when the signal is $\s$: 
\begin{align}\label{q_f}
    \fr(\s)=\qI_r(\s)+\qU, \quad \forall \r \in \R, \quad \forall \s \in \S. 
\end{align}

The loss function of the central authority is the average amount of traffic that exceeds a given threshold (traffic spillover) on one of the two routes. The authority may select the route(s) and threshold(s) for regulating traffic flows based on factors such as desirable or enforced capacity limits and/or maximum admissible flow through various routes, to limit direct impact (e.g., congestion) or indirect impact (e.g., noise or safety concerns) of traffic flow. For ease of exposition, we define the loss function as the average spillover on $\rtwo$ given a fixed threshold $\tau$ on the route flow: 
\begin{align}\label{loss_fun}
    L(\pi, \f)\deleq\sum_{\s \in \S} \P(\s) (f_r(\s)-\tau)_{+},
\end{align}
where $\P(\s)=\sum_{\w \in \W} \theta(\w)\pi(\s|\w)$ is the marginal probability with which the authority sends signal $\s$, and $(\f_2(\s)-\tau)_{+} = \max \{\f_2(\s)-\tau, 0\}$ is the amount of traffic that exceeds the threshold $\tau$ on $\rtwo$ when the signal is $\s$. Our subsequent analysis can also be applied to address the spillover on $\rone$ and/or arrive at a trade-off between desirable flows on $\rone$ and $\rtwo$. 

\subsection{Information Design Problem}
We formulate our information design problem as a two stage game: In the first stage, the central authority chooses the information structure $\pi$, and sends the realized signal $\s$ to all population $\I$ players. The signal creates a heterogeneous information environment among players since $\s$ is the private knowledge (type) of population $\I$. In the second stage, players play a Bayesian routing game under the information environment created by $\pi$ and $\s$. 

Given any information structure $\pi$, we denote the subgame in the second stage as $G(\pi)$. The \emph{common prior} of $G(\pi)$ is the joint probability of the state and the signal, denoted $\mu=\left(\mu(\w, \s)\right)_{\w \in \W,\s \in \S}$: 
\begin{align}\label{common_prior}
    \mu(\w, \s)= \theta(\w) \pi(\s|\w), \quad \forall \w \in \W, \quad \forall \s \in \S. 
\end{align}
The belief of population $\I$ given the received signal $\s$ is $\beta^{\s}=\left(\beta^{\s}(\w)\right)_{\w \in \W}$, where for any $\w \in \W$ and any $\s \in \S$, $\beta^{\s}(\w)$ is the posterior obtained from Bayes' rule: 
\begin{align}\label{belief_I}
    \betaIs(\w)=\frac{\theta(\w) \pi(\s|\w)}{\theta(\a)\pi(\s|\a)+\theta(\n) \pi(\s|\n)}= \frac{\mu(\w, \s)}{\P(\s)}.
\end{align}
Since $\pi$ satisfies \eqref{informativeness}, we can check that $\betaIa(\a) \geq \betaIn(\a)$ and $\betaIn(\n) \geq \betaIa(\n)$, i.e. the belief of state $\w$ when receiving signal $\s=\w$ is higher than that with the other signal. 

Given any feasible routing strategy profile $\q \in \Q$, the expected cost of each route $\r \in \R$ based on $\betaIs$ is as follows: 
\begin{align*}
    \mathbb{E}^1[c_r(\q)|\s]&=\sum_{\w \in \W} \betaIs(\w) c_r^{\w}(\fr(\s)), \quad \forall \s \in \S, \text{} \forall \r \in \R,
\end{align*}
where $\f$ is the aggregate route flow as in \eqref{q_f}.

Since population $\U$ does not have private information of the state, the expected cost of each route $\r$ is obtained based on the common prior $\mu$: 
\begin{align*}
\mathbb{E}^2[c_r(\q)]&=\sum_{\w \in \W}\sum_{\s \in \S} \mu(\w, \s) c_r^{\w}(\fr(\s)), \quad \forall \r \in \R.
\end{align*}
For the subgame $G(\pi)$, a routing strategy profile $\qwe$ is a Bayesian Wardrop equilibrium if for any $\r, \r' \in \R$, and any $\s \in \S$:
\begin{subequations}\label{q_eq}
\begin{align}
  \qIwe_r(\s)>0 \text{ } \Rightarrow \text{ } & \mathbb{E}^1[c_r(\qwe)|\s] \leq \mathbb{E}^1[c_{r'}(\qwe)|\s],\\
  \qUwe_r>0 \text{ } \Rightarrow \text{ } & \mathbb{E}^2[c_r(\qwe)] \leq \mathbb{E}^2[c_{r'}(\qwe)].
\end{align}
\end{subequations}
A route flow vector $\fwe$ is an equilibrium route flow if it can be induced by an equilibrium strategy profile $\qwe$. We do not explicitly indicate the dependence of $\qwe$ and $\fwe$ on $\pi$ for notational convenience.






The design of optimal information structure $\piwe$ can be formulated as the following optimization problem:
\begin{align}
    \min_{\pi} \quad &L(\pi, \fwe) \notag\\
    s.t. \quad &\text{$\pi$ satisfies \eqref{feasible_pi}}, \label{ecall}\\
    & \text{$\fwe$ is an equilibrium route flow vector in $G(\pi)$.} \notag
\end{align}

Formally, the tuple $(\piwe, \qwe)$ is a subgame perfect equilibrium (SPE) of the two-stage game if: 
\begin{itemize}
    \item[-] Given any information structure $\pi$, the routing strategy $\qwe$ is a Bayesian Wardrop equilibrium of the subgame $G(\pi)$, i.e. $\qwe$ satisfies \eqref{q_eq}.
    \item[-] The optimal information structure $\piwe$ minimizes the average traffic spillover given the correponding equilibrium route flow, i.e. $\piwe$ is the optimal solution of \eqref{ecall}.
    \end{itemize}

%

\section{Bayesian Wardrop Equilibrium}\label{sec:BWE}
To solve for the optimal information structure, we first need to analyze players' equilibrium routing decisions in any subgame $G(\pi)$. 

The following result parametrically characterizes the unique equilibrium route flow vector in $G(\pi)$ for any $\lamb \in [0,1]$. In this result, we show that given any information structure $\pi$, the properties of $\fwe$ depends on the relative size between $\lamb$ and the value of a function $g: \pi \to [0,1]$ defined as follows: 
\begin{align}\label{threshold}
    \lambbar(\pi)\deleq \frac{\atwo\D+\btwo-\bone}{\left(\bar{\aone}(\betaIn)+\atwo\right)\D}- \frac{\atwo\D+\btwo-\bone}{\left(\bar{\aone}(\betaIa)+\atwo\right)\D},
\end{align}where $\bar{\aone}(\beta^{\s})= \aonea \betaIs(\a)+\aonen \betaIs(\n)$, and the beliefs $\left(\betaIs(\w)\right)_{\w \in \W}$ are calculated as in \eqref{belief_I}. Since we consider two routes, and the central authority aims at minimizing the traffic spillover on $\rtwo$, we only present the equilibrium route flow on $\rtwo$. The flow on $\rone$ is $\fwe_1(\s)=\D-\fwe_2(\s)$ for any $\s \in \S$. 

\begin{proposition}\label{prop:BWE}
For any $\pi$ and $\lamb$, the equilibrium route flow is unique and satisfies the following properties:
\begin{itemize}
    \item[-] [$\lambbar(\pi) \geq \lamb$.] Population $\I$ exclusively takes $\rone$ when they receive signal $\n$, and $\rtwo$ with signal $\a$; population $\U$ splits on the two routes. The flow on $\rtwo$ is:

    {\footnotesize\begin{subequations}\label{lamb_small_eq}\begin{align}
    \fwe_2(\n)&= \D-\frac{\atwo\D+\btwo-\bone+\lamb\D \pro(\a) \left(\bar{\aone}(\betaIa)+\atwo\right)}{\bar{\aone}(\theta)+\atwo} \\
    \fwe_2(\a)&= \fwe_2(\n) + \lamb\D,
\end{align}
\end{subequations}}where  $\bar{\aone}(\theta)=p \aonea+(1-p) \aonen$. 
\item[-] [$\lambbar(\pi)<\lamb$.] Both populations split on the two routes. The flow on $\rtwo$ is: 
\begin{subequations}\label{large_lamb_eq}
\begin{align}  \fwe_2(\n)&=\D-\frac{\atwo \D+\btwo-\bone}{\bar{\aone}(\betaIn)+\atwo},\\
    \fwe_2(\a)&=\D-\frac{\atwo \D+\btwo-\bone}{\bar{\aone}(\betaIa)+\atwo}.
\end{align}
\end{subequations}
\end{itemize} 
\end{proposition}
\vspace{0.1cm}
The idea of the proof follows Theorem 1 of \cite{wu2017informational} and Theorem 1-2 in \cite{wu2018value}.

We now discuss the impact of information structure on the equilibrium route flows. Given any fraction $\lamb$, the set of feasible information structures can be partitioned into two sets as follows: 
\begin{align}\label{Pione_two}
    \Pi^1 \deleq \left\{\Pi|g(\pi) \geq \lamb \right\}, \quad \Pi^2 \deleq \left\{\Pi| g(\pi) < \lamb \right\}.
\end{align}
Proposition \ref{prop:BWE} shows that the impact of $\pi$ on $\fwe$ for the case when $\pi \in \Pione$ is distinct from $\pi \in \Pitwo$: 

For $\pi \in \Pione$, all players in population $\I$ deviate from choosing $\rone$ to $\rtwo$ when the received signal changes from $\n$ to $\a$. From \eqref{lamb_small_eq}, the change of flow on $\rtwo$ induced by the change of signal is $\fwe_2(\a)-\fwe_2(\n)=\lamb\D$, which does not depend on the information structure $\pi$. Moreover, as $\lamb$ increases, $\fwe_2(\n)$ decreases and $\fwe_2(\a)$ increases.

For $\pi \in \Pitwo$, both populations split on the two routes. From \eqref{large_lamb_eq}, the change of flow on $\rtwo$ induced by the change of signal is $\fwe_2(\a)-\fwe_2(\n)=\lambbar(\pi)$. The value of $\lambbar(\pi)$ in \eqref{threshold} increases in $\betaIa(\a)-\betaIn(\a)$, which evaluates the relative difference between the beliefs of state $\a$ given the two signals. 

Furthermore, for any $\pi \in \Pitwo$, the equilibrium route flows in \eqref{large_lamb_eq} do not change with $\lamb$. This implies that any equilibrium outcome when only $\lamb$ fraction of players receive signal $\s$ according to $\pi \in \Pitwo$ is equivalent to the case where \emph{all} players receive the signal. This property will be used in Sec. \ref{sec:persuasion} for identifying an interval of $\lamb$, for which the optimal information structure and the equilibrium outcome do not depend on $\lamb$.

%
%

Finally, note that the information structure $\pi$ affects the value of $\lambbar(\pi)$ in \eqref{threshold}, and the equilibrium route flows in \eqref{lamb_small_eq}
 - \eqref{large_lamb_eq} through the beliefs and the probability of signals defined as follows:
 \begin{align*}
 \beta \deleq (\betaIa, \betaIn), \quad \P \deleq (\pro(\a), \pro(\n)).
 \end{align*}
Then, we can re-write $L(\pi, \fwe)$ in \eqref{loss_fun} as a function of $(\beta, \P)$, denoted $\bar{L}(\beta, \P)$, and $\lambbar(\pi)$ as a a function of $\beta$, denoted $\gbar(\beta)$. The characterization of how the equilibrium route flow depends on $(\beta, \P)$ and the fraction $\lamb$ is crucial for our approach for solving the optimal information design problem in the next section. 
\section{Optimal Information Design}\label{sec:persuasion}
In this section, we present the optimal information structure $\piwe$, and analyze how $\piwe$ changes with the fraction $\lamb$ and the flow threshold $\tau$. 

Due to the space limit, we only present the optimal information structure for cases where the threshold $\tau$ satisfies the following constraint:

{\footnotesize \begin{align}\label{threshold_range}
\D-\frac{\atwo\D+\btwo-\bone}{\aonen+\atwo} \leq \tau \leq \D-\frac{\atwo\D+\btwo-\bone}{\aonea+\atwo}.
\end{align}}The lower (resp. upper) bound of $\tau$ is the equilibrium route flow on $\rtwo$ when all players have complete information of the state $\n$ (resp. $\a$). Therefore, \eqref{threshold_range} means that in complete information environment, the spillover is positive in state $\a$, but zero in state $\n$. Our solution approach can be easily extended to the cases where $\tau$ is outside of this range. 


We first obtain that if the prior probability of state $\a$ is low, then the optimal information structure is to provide no information of the state. 
\begin{proposition}\label{prop:no_persuasion}
If $p \leq \bar{p}$, where 
\begin{align*}
    \bar{p} = \frac{1}{\aonea-\aonen} \left(\frac{\atwo\D+\btwo-\bone}{\D-\tau} - \atwo -\aonen\right), 
\end{align*}
then the optimal information structure is to provide no information of the state, i.e. $\piwe(\s|\w)=\P(\s)$ for any $\s \in \S$ and $\w \in \W$. The average traffic spillover is $L(\piwe, \fwe)=0$. 
\end{proposition}
From Proposition \ref{prop:BWE}, we know that if $\piwe$ does not provide state information, then $\lambbar(\piwe)=0$, and the equilibrium route flow is as follows: 
\begin{align}\label{eq_homo}
    \fwe_2(\s)=\D-\frac{\atwo \D+
    \btwo-\bone}{\bar{\aone}(\theta)+\atwo}, \quad \forall \s \in \S.
\end{align}
The value of $\pbar$ is the threshold of $p$ such that $\fwe_2(\s)$ in \eqref{eq_homo} equals to $\tau$. For any $p\leq \pbar$, $\fwe_2(\s) \leq \tau$. Therefore, the objective function in \eqref{ecall} is zero (attains the minimum) when the information structure provides no information of the state. From \eqref{threshold_range}, we know that $\pbar \in [0,1]$.

For $p>\pbar$, based on Proposition \ref{prop:BWE}, we can restate the optimal information design problem \eqref{ecall} as the following optimization problem:   
 \begin{equation}\label{opt_basic}
 \begin{split}
     \min_{\pi} \quad &L(\pi, \fwe)=\sum_{\s \in \S}\pro(\s)  (\fwe_2(\s)-\tau)_{+}, \\
     s.t. \quad & 
     \begin{array}{ll}
     \text{$\fwe_2$ is in \eqref{lamb_small_eq}}, & \text{if $ \lambbar(\pi)\geq\lamb$,}\\
     \text{$\fwe_2$ is in \eqref{large_lamb_eq}}, & \text{if $\lambbar(\pi) < \lamb$},
     \end{array}
     \\
     & \text{ $\pi$ satisfies \eqref{feasible_pi},}
 \end{split}
 \end{equation}
 where $\lambbar(\pi)$ is given by \eqref{threshold}.

The optimization problem \eqref{opt_basic} is non-linear and non-convex in the information structure $\pi$. The key difficulties in solving \eqref{opt_basic} are: (i) the value of $\lambbar(\pi)$, and the equilibrium route flows $\fwe_2$ are nonlinear functions of $(\beta, \P)$, which are again nonlinear functions of $\pi$; (ii) the expressions of the equilibrium route flows are different for information structures in $\Pione$ and $\Pitwo$; (iii) the objective function is a piece-wise linear (instead of linear) function of the equilibrium route flows.

We develop an approach to tackle these difficulties and solve the optimal information structure analytically: First, we characterize the set of $(\beta, \P)$ induced by information structure $\pi$ satisfying \eqref{feasible_pi}, which can be used to construct another optimization problem to solve the optimal $(\beta^{*}, \Pwe)$ directly (Lemma \ref{lemma_one}). Second, we identify the range of $\lamb$ in which the optimal information structure satisfies $\piwe \in \Pione$, and the equilibrium route flow is given by \eqref{lamb_small_eq} (Lemma \ref{lemma_two}). Third, we prove that the equilibrium flow on $\rtwo$ under optimal information structure is no less than $\tau$; thus $L(\pi, \fwe)$ is equivalent to a linear function of $\fwe$ (Lemma \ref{lemma_three}).

\begin{lemma}\label{lemma_one}
A tuple $(\beta, \P)$ can be induced by a feasible information structure $\pi$ if and only if $(\beta, \P)$ satisfies: 
\begin{subequations}\label{Bayesian_plausible}
\begin{align}
    &\betaIa(\a) \cdot \pro(\a) + \betaIn(\a) \cdot \pro(\n)  = p, \label{eq_plausible}\\
    &\betaIa(\a) \geq \betaIn(\a),\label{beta_order}\\
    &\betaIa(\n)+\betaIa(\a)=1, \betaIa(\a), \betaIa(\n) \geq 0, \label{feasible_one}\\  &\betaIn(\n)+\betaIn(\a)=1, \betaIn(\n),\betaIn(\a) \geq 0, \\
    &\pro(\a)+\pro(\n)=1,\quad  \pro(\a), \pro(\n) \geq 0. \label{feasible_two}
\end{align}
\end{subequations}
 
\end{lemma}
The idea of the proof follows Proposition 1 in \cite{kamenica2011bayesian}. 

Constraint \eqref{eq_plausible} ensures that $\beta$ is derived from $\theta$ and $\pi$ as in \eqref{belief_I}. Constraint \eqref{beta_order} results from \eqref{informativeness} to exclude beliefs that are induced by equivalent information structures. Constraints \eqref{feasible_one} -- \eqref{feasible_two} ensure that $\beta$ and $\P$ are feasible probability vectors. 

Based on Lemma \ref{lemma_one} and following \eqref{opt_basic}, we can solve for the optimal $(\beta^{*}, \Pwe)$ from the following optimization problem: 
 \begin{equation}\label{opt_beta}
 \begin{split}
     \min_{\beta, \P} \quad &\bar{L}(\beta, \P)=\sum_{\s \in \S}\pro(\s)  (\fwe_2(\s)-\tau)_{+}, \\
     s.t. \quad & 
     \begin{array}{ll}
     \text{$\fwe_2$ is in \eqref{lamb_small_eq}}, & \text{if $ \gbar(\beta)\geq\lamb$,}\\
     \text{$\fwe_2$ is in \eqref{large_lamb_eq}}, & \text{if $\gbar(\beta) < \lamb$},
     \end{array}
     \\
     & \text{ $(\beta, \P)$ satisfies \eqref{Bayesian_plausible},}
 \end{split}
 \end{equation}
 where $\gbar(\beta)$ is a function of $\beta$ as in \eqref{threshold}. We use both \eqref{opt_basic} and \eqref{opt_beta} for designing the optimal information structure.
 
Next, we identify a threshold $\lambddag \in (0,1)$ as follows: 
\begin{align}\label{lambddag}
    \lambddag &= 1- \frac{\atwo\D+\btwo-\bone}{\(\aonea+\atwo\)\D}-\frac{\tau}{\D}.
\end{align}
\begin{lemma}\label{lemma_two}
For any $p> \pbar$, and any $\lamb < \lambddag$, the optimal information structure $\piwe \in \Pione$, i.e. $\lambbar(\piwe)\geq\lamb$, where $\lambbar(\piwe)$ is in \eqref{threshold}. The equilibrium route flow is given by \eqref{lamb_small_eq}.  
\end{lemma}


Furthermore, we show that given the optimal information structure, $\fwe_2(\s)$ is no less than the threshold $\tau$ for any $\s \in \S$. 
%
\begin{lemma}\label{lemma_three}
For any $p>\pbar$ and any $\lamb \in [0, 1]$, the equilibrium route flows induced by the optimal information structure must satisfy $\fwe_2(\s) \geq \tau$ for any $\s \in \S$.  
\end{lemma}

Lemma \ref{lemma_three} shows that the objective function in \eqref{opt_basic} can be simplified as a linear function of $\fwe$:
\begin{align}
    L(\pi, \fwe)&=\P(\a) (\fwe_2(\a)-\tau)+ \P(\n)(\fwe_2(\n)-\tau)\notag\\
    &=\P(\a)\fwe_2(\a)+ \P(\n)\fwe_2(\n)-\tau.
\end{align}

We are now ready to derive the optimal information structure $\piwe$. We find that $\piwe$ is different for $\lamb$ in three regimes: $\Lambone: \lamb \in [0, \left. \lambdag\right)$; $\Lambtwo: \lamb \in [\lambdag, \lambddag)$, and $\Lambthree: \lambda \in [\lambddag, 1]$. The threshold $\lambdag$ is given by:
\begin{align}\label{lambdag}
    \lambdag& =\frac{\left(\D-\tau\right) \left(\bar{\aone}(\theta)+\atwo\right)-\atwo \D -\btwo +\bone}{\D p (\aonea+\atwo)},
\end{align}
and $\lambddag$ is given by \eqref{lambddag}. Since $\tau$ satisfies \eqref{threshold_range} and $p>\bar{p}$, we can check that $0<\lambdag < \lambddag <1$. Therefore, the three regimes are well-defined intervals of $\lamb$. 

Based on Proposition \ref{prop:BWE} and Lemmas \ref{lemma_one} -- \ref{lemma_three}, we characterize the optimal information structure in each regime. We also present the equilibrium route flow and the average traffic spillover in each regime. 
\begin{theorem}\label{theorem:opt_persuasion}
For any $p>\bar{p}$, \\
\noindent In regime $\Lambone$, the optimal information structure is:
\begin{subequations}\label{one_persuasion}
\begin{align}
    \piwe(\a|\n)&=0, \quad \piwe(\n|\n)=1, \\
    \piwe(\a|\a)&=1, \quad 
    \piwe(\n|\a)=0.
\end{align}
\end{subequations}
The equilibrium route flow is: 
\begin{align*}
        \fwe_2(\n)&=\D-\frac{\atwo\D+\btwo-\bone+\lamb\D p(\aonea+\atwo)}{\bar{\aone}(\theta)+\atwo}\\
        \fwe_2(\a)&= \fwe_2(\n)+\lamb\D.
    \end{align*}
    The average traffic spillover decreases in $\lamb$:
    {\footnotesize \begin{align*}
        L(\piwe, \fwe)=\D-\tau-\frac{\atwo\D+\btwo-\bone}{\bar{\aone}(\theta)+\atwo} -\frac{p(1-p)(\aonea-\aonen)\lamb \D}{\bar{\aone}(\theta)+\atwo}.
    \end{align*}}
\noindent In regime $\Lambtwo$, the optimal information structure is: 

\begin{subequations}\label{two_persuasion}
\begin{align}
    \piwe(\a|\n)&=0, \quad \piwe(\n|\n)=1, \\
    \piwe(\a|\a)&=\frac{(\D-\tau)(\bar{\aone}(\theta)+\atwo)-\atwo\D-\btwo+\bone}{\lamb \D (\aonea+\atwo) p}, \label{piaa_two}\\
    \piwe(\n|\a)&=1-\piwe(\a|\a).
\end{align}
\end{subequations}
The equilibrium route flow is: 
    \begin{align*}
        \fwe_2(\n) &= \tau, \quad
        \fwe_2(\a) = \tau+\lamb\D.
    \end{align*}
    The average traffic spillover does not change with $\lamb$:
    {\footnotesize \begin{align}\label{c_two}
        L(\piwe, \fwe)=\frac{(\D-\tau)(\bar{\aone}(\theta)+\atwo)-\atwo\D-\btwo+\bone}{\aonea+\atwo}.
    \end{align}}
In regime $\Lambthree$, the optimal information structure is: 
\begin{subequations}\label{three_persuasion}
\begin{align}
    \piwe(\a|\n)&=0, \quad \piwe(\n|\n)=1, \\
    \piwe(\a|\a)&=\frac{(\D-\tau)(\bar{\aone}(\theta)+\atwo)-\atwo\D-\btwo+\bone}{\left((\D-\tau)(\aonea+\atwo)-\atwo\D-\btwo+\bone\right)p}, \\
    \piwe(\n|\a)&=1-\piwe(\a|\a).
\end{align}
\end{subequations}
The equilibrium route flow is: 
\begin{subequations}\label{pi_three}
    \begin{align}
        \fwe_2(\n)&=\tau, \quad
        \fwe_2(\a)=\D-\frac{\atwo\D+\btwo-\bone}{\aonea+\atwo}.
    \end{align}
    \end{subequations}
    The average traffic spillover $L(\piwe, \fwe)$ is as in \eqref{c_two}.  
\end{theorem}


Now we discuss the properties of optimal information structure in detail. Firstly, in state $\n$, the signal provides complete state information, i.e. $\piwe(\n|\n)=1$. This is because when the cost on $\rone$ is low in state $\n$, sending signal $\a$ will unnecessarily increase the traffic spillover on $\rtwo$. In state $\a$, the signal provides complete state information when the fraction of population 1 is smaller than $\lambdag$, but only provides partial state information ($\piwe(\a|\a)<1$) if $\lamb>\lambdag$ to avoid sending large flow to $\rtwo$.

Secondly, the average spillover decreases with $\lamb$ in regime $\Lambone$ ($\lamb < \lambdag$), and does not change with $\lamb$ in regimes $\Lambtwo$ and $\Lambthree$ ($\lamb \geq \lambdag$). This implies that the minimum average traffic spillover can be achieved by the optimal information structure as long as the fraction of travelers receiving the signal exceeds the threshold $\lambdag$, which is smaller than 1. Moreover, if $\lamb \geq \lambdag$, then the spillover is only positive in state $\a$, i.e. traffic flow on $\rtwo$ only exceeds the threshold flow $\tau$ if there is an incident.  The probability of positive spillover with optimal information structure ($\mu^{*}(\a, \a)=\piwe(\a|\a) \cdot p$) is smaller than that in the case where players have no state information (the spillover probability is 1) and the case where players have complete state information (the spillover probability is $p$). 

Thirdly, $\lambddag$ is the threshold fraction beyond which the optimal information structure $\piwe$ and the equilibrium route flow $\fwe$ do not depend on $\lamb$. Additionally, $\lambddag$ is the maximum impact of the signal on route flows, i.e. the maximum fraction of travelers who change routing decisions with the received signals. For any $\lamb \leq \lambddag$, $\piwe \in \Pione$ and the signal influences the routing decisions of all travelers in population 1 ($\lamb< \lambddag$ fraction). On the other hand, for any $\lamb > \lambddag$, $\piwe \in \Pitwo$. Then, regardless of the fraction of population 1, $\lambddag$ fraction of travelers change their routing decisions with the signal.

Finally, we find that the regime boundaries $\lambdag$ and $\lambddag$ in \eqref{lambdag} and \eqref{lambddag} decrease as $\tau$ increases. Practically, this implies that if more traffic can be routed on $\rtwo$ (i.e. $\tau$ is larger), then the minimum average spillover can be achieved by sending signals to a smaller fraction of players ($\lambdag$) according to the optimal information structure. Additionally, the maximum fraction of players who change routing decisions with the received signals ($\lambddag$) is also smaller. 

\section{Impact of information design on travel costs}\label{sec:cost}
We now analyze how the optimal information structure designed for minimizing the spillover affects the players' cost (both individual and social) in equilibrium. 

We define equilibrium population costs as the average travel time costs experienced by players in each population in equilibrium: $C^{\I*}= \frac{1}{\lamb \D} \sum_{\w \in \W} \sum_{\s \in \S} c_r^{\w}(\fwe_r(\s))\cdot q_r^{\I*}(\s)$ and $C^{\U*}= \frac{1}{(1-\lamb) \D} \sum_{\w \in \W} \sum_{\s \in \S} c_r^{\w}(\fwe_r(\s))\cdot q_r^{\U*}$.
We define the equilibrium average cost of all players as $C^{*}=\lamb C^{\I*} + (1-\lamb) C^{\U*}$.

For any $p\leq\pbar$, we know from Proposition \ref{prop:no_persuasion} that the optimal information structure provides no state information to players. Hence, the information structure has no impact on the players' equilibrium costs. 

For any $p > \pbar$, we can compute the equilibrium population costs and the equilibrium average cost based on Proposition \ref{prop:BWE} and Theorem \ref{theorem:opt_persuasion}.
\begin{proposition}\label{eq_route_cost}
Given the optimal information structure $\piwe$, $C^{\I*} \leq C^{\U*}$ if $\lamb \in [0,\lambddag)$, and $C^{\I*} = C^{\U*}$ if $\lamb \in [\lambddag,1]$. 
Furthermore, as $\lamb$ increases, $C^{*}$ monotonically decreases in regime $\Lambone$, increases in regime $\Lambtwo$, does not change in regime $\Lambthree$. \hfill$\blacksquare$ 
\end{proposition} 

\vspace{0.1cm}
Proposition \ref{eq_route_cost} shows that the signal sent by the central authority gives players in population $\I$ an advantage over population $\U$ in terms of the average costs if $\lamb < \lambddag$, and the two populations experience the same cost if $\lamb \geq \lambddag$. Furthermore, Theorem \ref{theorem:opt_persuasion} and Proposition \ref{eq_route_cost} show that in regime $\Lambone$, increasing $\lamb$ reduces both the traffic spillover on $\rtwo$ and the equilibrium average cost. However, if $\lamb$ increases beyond $\lambdag$, then the average cost increases, while the average spillover does not change. 

One can interpret these insights in the context of two practical situations -- the fraction $\lamb$ is induced by players' choice of accessing to the signal versus chosen by the designer versus. In both situations, the average traffic spillover is the same. 

If players can choose whether or not to get access to signals -- the information, then the fraction of population $1$ will be higher or equal to $\lambddag$. This is because players in population 2 will continue to switch to population 1 until the costs of two populations are the same. Hence, the optimal information structure $\piwe$ is given by \eqref{three_persuasion}. 

On the other hand, if the designer can choose the fraction $\lamb$ as well as the information structure $\pi$, then it is optimal for the authority to provide complete state information as in \eqref{one_persuasion} to $\lambdag$ fraction of players in order to achieve the minimum average spillover and the minimum average cost. However, in this case, $C^{1*}<C^{2*}$. Therefore, the resulting information structure favors the set of players who receive the signals. 


We illustrate our results in the following example.
\begin{example}
The cost functions of the network are $c_1^{\a}(f_1)=3 \f_1 + 15$, $c_1^n(f_1)=f_1+15$, and $c_2(f_2)=2 f_2 +20$. The total demand $\D=10$, and the threshold $\tau=2.5$. The probability of state $\a$ is $p=0.3$. From \eqref{lambdag} and \eqref{lambddag}, the thresholds are $\lambdag=0.133$, and $\lambddag=0.25$. 

Fig. \ref{fig:provision} shows $\piwe(\a|\a)$ for $\lamb \in [0,1]$. Fig. \ref{popu_plot} shows the resulting equilibrium cost of each population. Fig. \ref{Lfigure} and Fig. \ref{Cplot} compare the average traffic spillover and the equilibrium average cost under the optimal information structure with the corresponding costs in two situations: (1) the central authority provides no information to players; (2)  complete information of the state is provided to $\lamb$ fraction of players. 

In this example, the minimum average spillover can be achieved as long as more than $13.3\%$ of players have access to the signal (Fig. \ref{Lfigure}). If the fraction increases over $25\%$, then the optimal information design does not depend on $\lamb$ (Fig. \ref{fig:provision}). Moreover, the optimal information structure achieves $18\%$ lower spillover in comparison to the case of no information, and $47\%$ lower spillover in comparison to providing complete state information to all players. This demonstrates that the central authority achieves non-trivial reduction in average spillover by optimal information design even when a high fraction of players do not have access to the information signal. 

Additionally, Fig. \ref{popu_plot} shows that in this example population 1 enjoys reduction of cost by $7\%$ compared with population 2 if $\lamb=\lambdag$, which is the fraction that minimizes the average spillover and the average cost. Finally, from Fig. \eqref{Cplot}, we see that the equilibrium average cost under optimal information structure is lower than that in the case with no state information, but the minimum cost under optimal information structure (when $\lamb=\lambdag$) is slightly ($1\%$) higher than the minimum cost in the case where the signal provides complete state information. 
\begin{figure}[htp]
    \centering
     \begin{subfigure}{0.4\textwidth}
    \includegraphics[width=\textwidth]{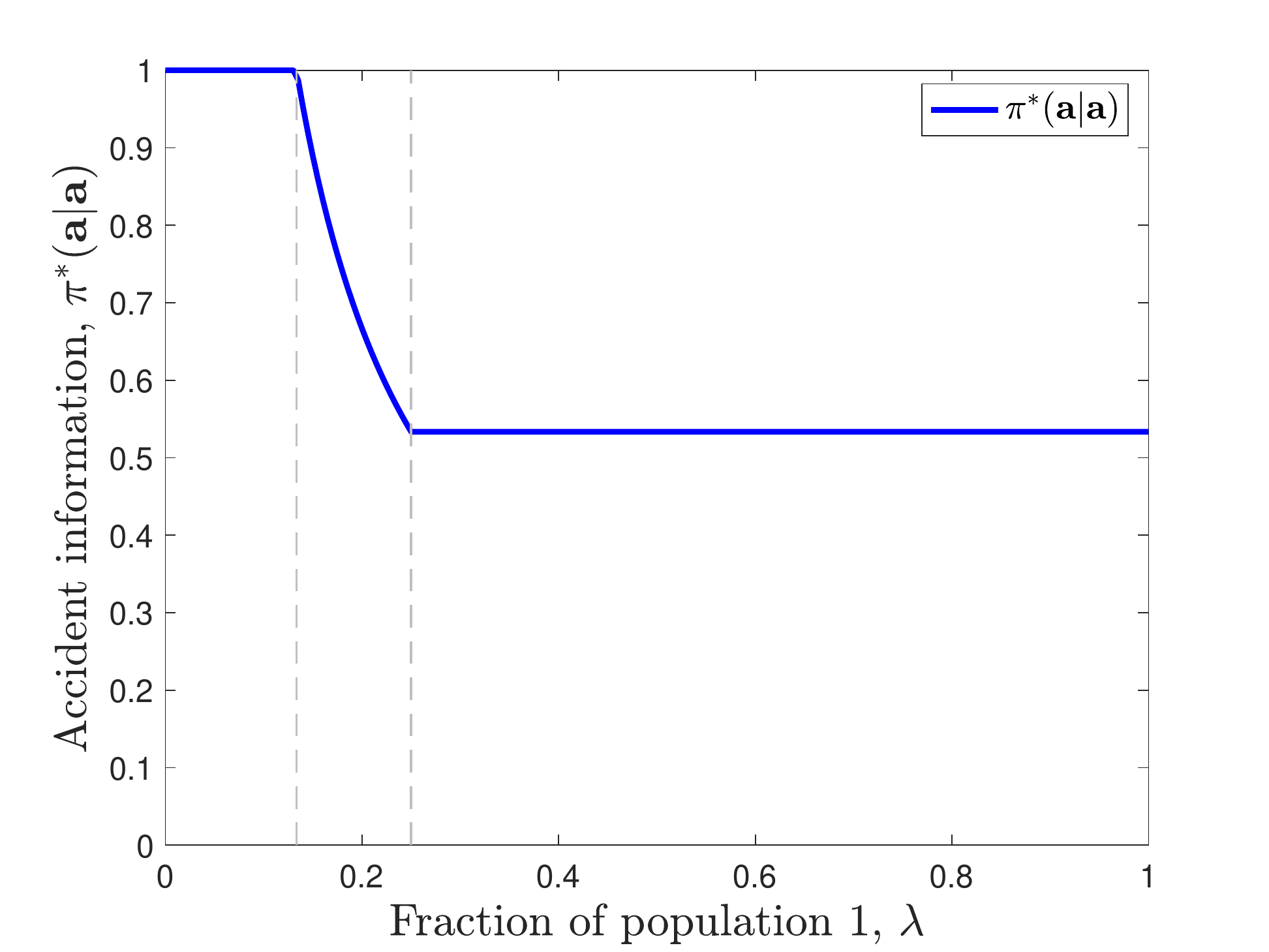}
    \caption{}
    \label{fig:provision}
    \end{subfigure}
    \begin{subfigure}{0.4\textwidth}
    \includegraphics[width=\textwidth]{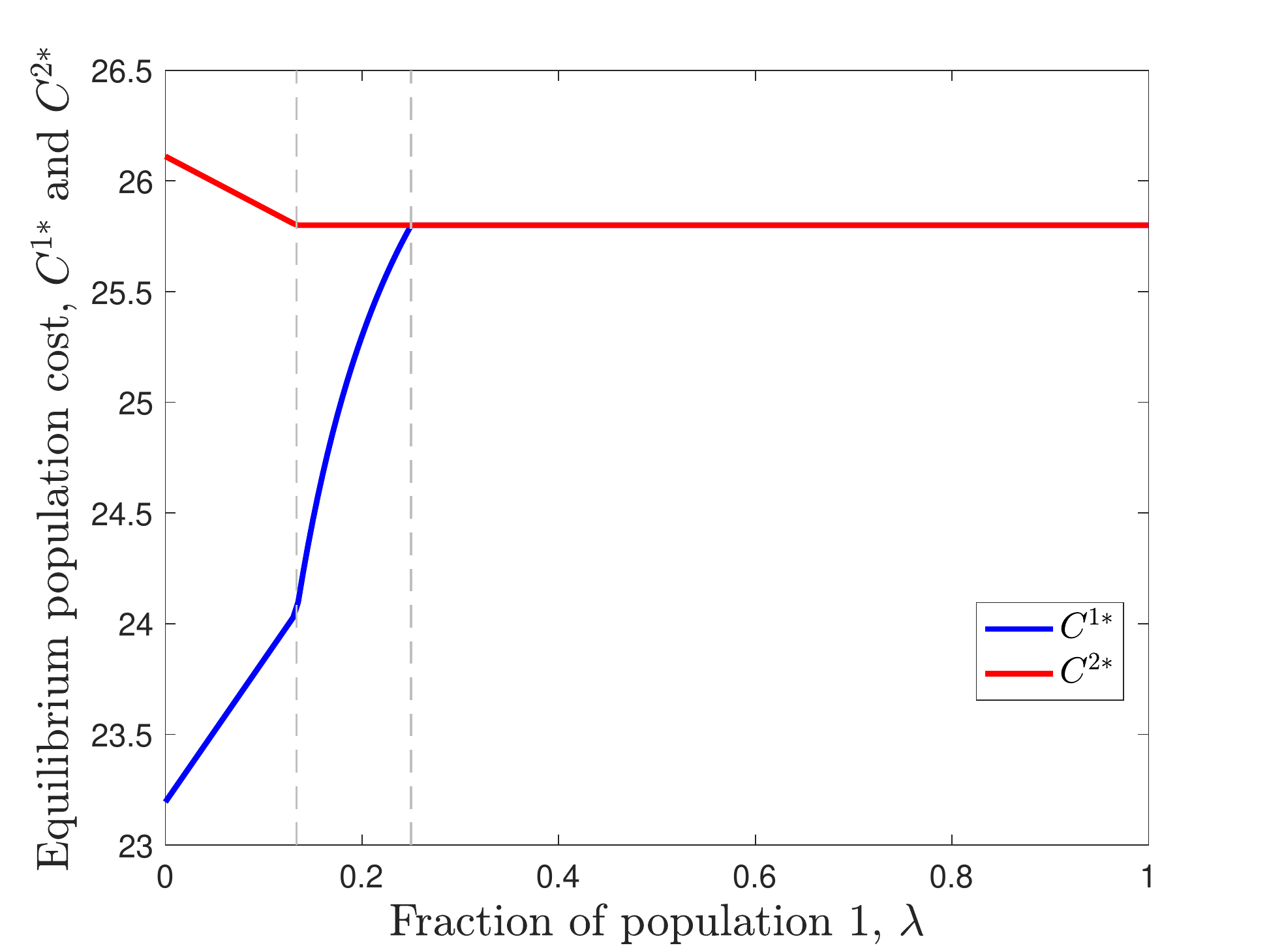}
    \caption{}
    \label{popu_plot}
    \end{subfigure}\\
    \begin{subfigure}{0.4\textwidth}
    \includegraphics[width=\textwidth]{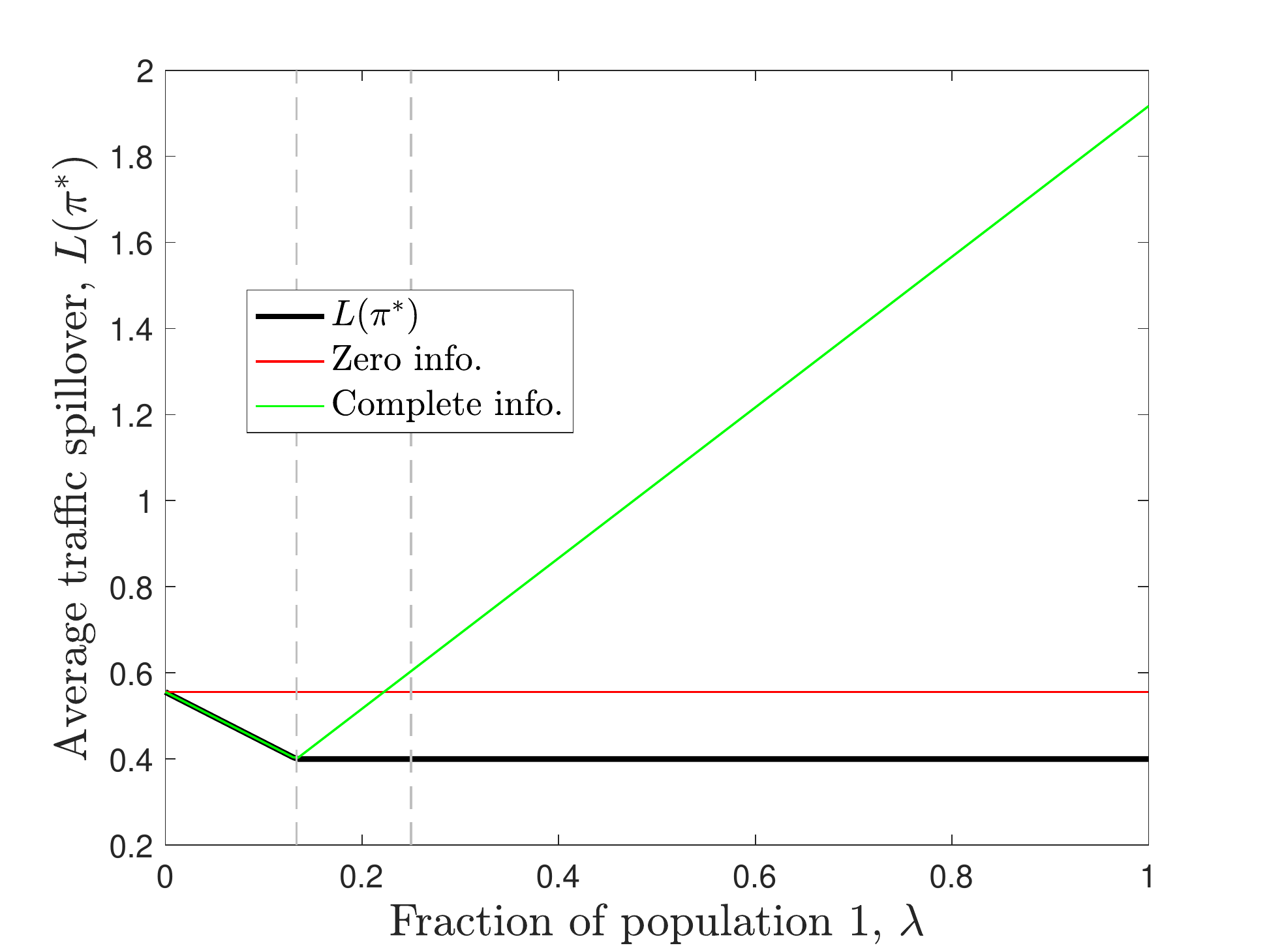}
    \caption{}
    \label{Lfigure}
    \end{subfigure}
    \begin{subfigure}{0.4\textwidth}
    \includegraphics[width=\textwidth]{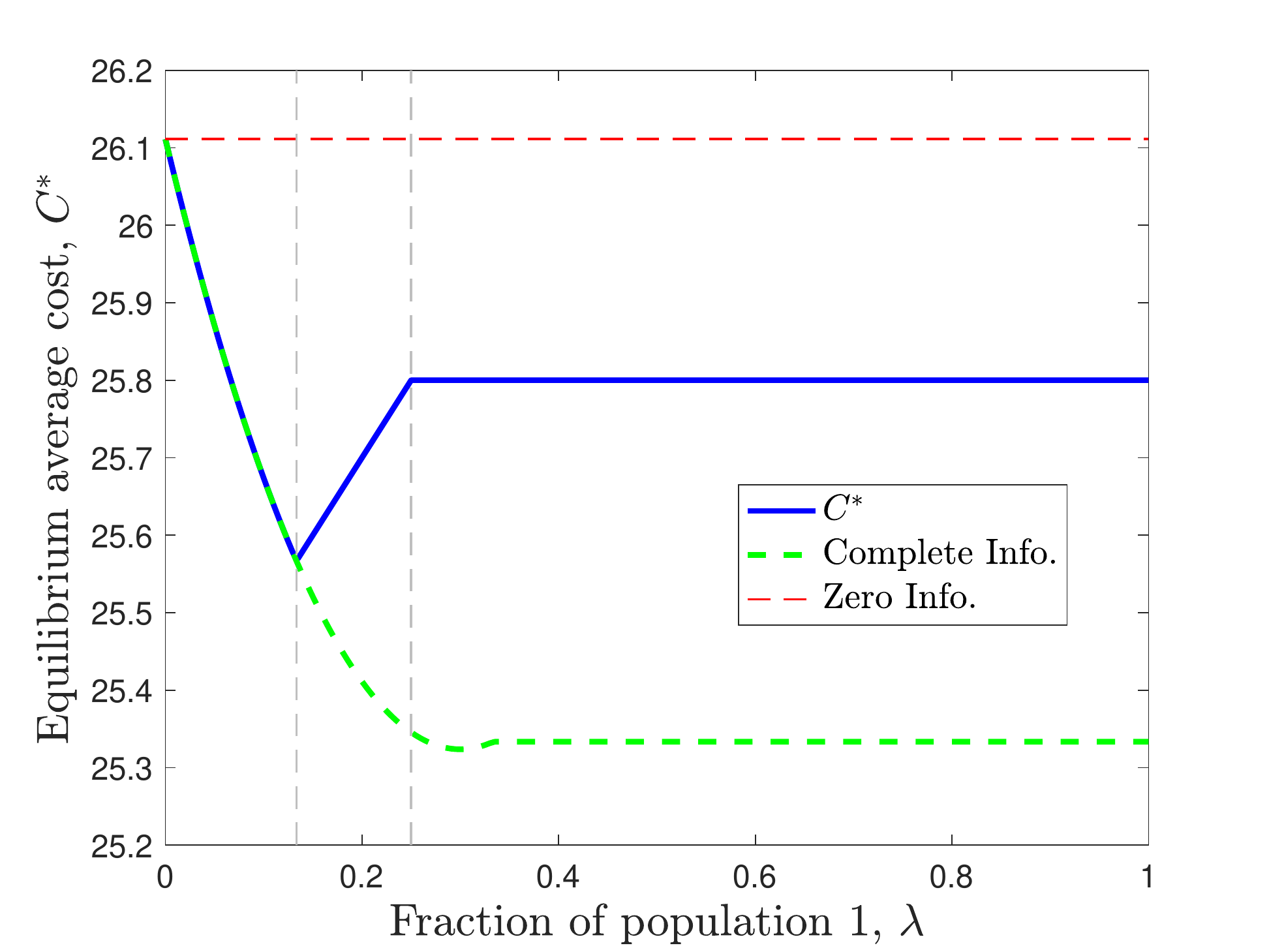}
    \caption{}
    \label{Cplot}
    \end{subfigure}
    \caption{Congestion costs under optimal information structure: (a) Probability of accident signal in state $\a$; (b) Equilibrium population costs; (c) Average spillover on $\rtwo$; (d) Equilibrium average cost.}
    \label{fig:my_label}
\end{figure}

\end{example}




\section{Concluding Remarks}
Our work studies the problem of information design for a central authority who aims to reduce traffic spillover on certain routes in a transportation network with uncertain states. Importantly, our model addresses the practical situation where only a fraction of players have access to the state information sent by the authority. We present an approach to analyze players' equilibrium routing strategies in heterogeneous information environment, and characterize the optimal information structure given any fraction of players receiving the signal. Our results suggest that the minimum spillover can be achieved by the optimal information design without ensuring that the signal is received by all players. We are interested in extending our results to networks with multiple routes and multiple states. We are also interested in investigating the tradeoff between regulating spillover and reducing average cost.

\section*{ACKNOWLEDGMENT}
We are grateful to Prof. Asu Ozdaglar and Prof. Demos Teneketzis for useful discussions. The first author was supported by the IDSS Hammer Fellowship, and the second author was party supported by NSF Grant No. 1239054 CPS Frontiers: Foundations Of Resilient CybEr-Physical Systems (FORCES), and NSF CAREER Award CNS 1453126.
\bibliographystyle{IEEE_tran}
\bibliography{library}


\end{document}